\documentclass[12pt]{article}
\begin{document}

\def \ep{\epsilon}
\def \intR {\int_{-\infty}^{+\infty}}
\def \grad {\nabla}
\def \ov{\over}
\def \q {\quad}
\def \qq {\qquad}
\def \bar{\overline}
\def \beq{\begin{equation} }
\def \eeq{\end{equation} }
\def \lb{\label}
\def \R{{\bf R}}

\newtheorem{lemma}{Lemma}
\newtheorem{theorem}{Theorem}
\newtheorem{proposition}{Proposition}
\newtheorem{definition}{Definition}

\baselineskip .55cm

\def \pd{\partial}
\def\~#1{\widetilde #1}
\def\.#1{\dot #1}
\def\^#1{\widehat #1}
\def\"#1{\ddot #1}
\def\d{{\rm d}}       
\def \id{\! :=}

\def \sy {symmetry}
\def \sys {symmetries}
\def \so {solution}
\def \eq{equation}
\def \hom {homoclinic }
\def \Ham {Hamiltonian}
\def\a{\alpha}
\def\be{\beta}
\def\phi{\varphi}
\def\de{\delta}
\def\om{\omega}
\def\la{\lambda}
\def\De{\Delta}
\def\th{\theta}
\def \qq{\qquad}
\def \q{\quad}
\def \pn{\paragraph\noindent}
\def \sk{\smallskip}
\def \ni{\noindent}

\title{{\bf Mel'nikov method revisited}}

\author{Giampaolo Cicogna\thanks{Email: cicogna@df.unipi.it} \\
Dip. di Fisica ``E.Fermi'' and I.N.F.N., Sez. di Pisa, \\
Via Buonarroti 2, Ed. B, I-56127, Pisa, Italy \\
and \\ Manuele Santoprete\thanks{Email: msantopr@math.uvic.ca} \\
Dept. of Mathematics and Statistics, University of Victoria, \\
P.O. Box 3045, Victoria B.C., Canada, V8W 3P4}

\date{}

\maketitle

\begin{abstract}
We illustrate a completely analytic approach to Mel'nikov theory,
which is based on a suitable extension of a classical method,
and which is parallel and -- at least in part -- complementary
to the standard procedure. This approach can be also
applied to some ``degenerate'' situations, as to the
case of nonhyperbolic unstable points, or of critical points
located at the infinity (thus giving rise to unbounded orbits, e.g. the
Keplerian parabolic orbits), and it is naturally ``compatible'' with
the presence of general symmetry properties of the problem.
These peculiarities may clearly make this approach of great interest
in celestial mechanics, as shown by some classical
examples.
\end{abstract}


\vfill\eject
~\vskip 1.2 truecm

\section{Introduction}
It is certainly impossible to give a fairly complete list of the papers
devoted to the applications of the Mel'nikov method \cite{Me,GH,W}
for evaluating the
onset of chaos arising in perturbed homoclinic (or heteroclinic) orbits.
We will quote only some of the papers which are more directly connected
with the present approach.

The main purpose of this paper is to illustrate a completely analytic
procedure, based on a suitable extension of a classical method
\cite{CHM,Gr}, which is parallel and -- at least in part -- complementary
to the standard approach (see e.g. \cite{GH,W,HM}). This procedure can be
also applied to some
``degenerate'' situations, as to the case of unstable nonhyperbolic
points, or of critical points located at the infinity (thus giving rise
to unbounded orbits, e.g. the Keplerian parabolic orbits), and it is
naturally ``compatible'' with the presence of general symmetry properties
of the problem. For these reasons, apart from a clear ``unifying'' aspect,
this method could be of great interest in celestial mechanics, and it could
be a contribution to the study of some of the questions about the onset of
chaos in unbounded phase space systems and in the presence of unbounded
orbits. It has been remarked indeed that in this situation chaos
manifests itself in a particularly dramatic way \cite{LM}.

Let us remark immediately that, in the above mentioned degenerate cases,
i.e. in the lack
of the hypothesis of hyperbolicity, standard results of perturbation
theory cannot be directly applied; for instance, to preserve the
criticality, we will have to impose a sufficiently
rapid vanishing of the perturbation at the critical point. We can then
extend the introduction of Mel'nikov functions, and show not only the
existence of smooth solutions of the perturbed problem, approaching the
critical points and playing in this context the role of
stable and unstable manifolds,
but also the possible presence of infinitely many intersections
of these asymptotic sets on the Poincar\'e sections, thus leading to a
complicate dynamics typical of the homoclinic chaotic behaviour
\cite{GH,W,HM,DH,CS2}.

Being mainly interested in the methodological aspects, we will not
devote special emphasis on new applications, but rather we will show,
in the two last sections of this paper, how
the method can be concretely applied in some typical situations, arising
especially in celestial mechanics and general relativity.

\bigskip

\section{Statement of the Problem}
Although the procedure is completely general (indeed, we will
state some results in a quite general setting in sections 3 and 4),
we have actually in mind  applications to celestial mechanics or to
gravitational problems in general relativity;
therefore we will restrict our attention mainly to problems described
by \Ham s of the classical form
\beq
H={1\ov 2}\Big(p_r^2+{p_\th^2\ov r^2}+{p_\phi^2\ov {r^2\sin^2\th}}\Big)
+V(r) \label{H0}
\eeq
where $V(r)$ is some ``long range'' potential. Concretely,
we have in mind potentials of the form
\beq
V(r)=-{1-\be_1\ov r}-{\be_2\ov r^2}-{\be_3\ov r^3} \label{pot}
\eeq
where $\be_i$ are given parameters, which model several interesting
situations, including standard Kepler problem, Kepler problems plus
quadrupolar effects \cite{LM}, Manev problem \cite{DS}, or the motion of
a charged particle in the field of a Schwarzschild black hole in
general relativity \cite{LV,SC}, and many other situations.

Let us write the \eq s of the motion, assuming for a moment that it lies
in the plane $\th=\pi/2$,
\beq
\.r=p_r \q ,\q \.p_r=-{dV\ov dr}+{L^2\ov r^3} \q , \q
\.\phi={L\ov r^2} \q , \q
\.p_\phi=0 \label{mot2}
\eeq
where we have put $L=p_\phi=\, $const$\ \not=0$. We are interested in this
paper in the appearance of chaotic behaviour related to the presence
of \hom (or heteroclinic)
orbits subjected to perturbations; therefore, the relevant situations
which may occur, depending on the values of the parameters $\beta_i$ in the
potential (\ref{pot}), and which we are going to deal with in this
paper, are the following:

\medskip\ni
{\em i)} the presence of an unstable equilibrium point $r_u\not=0$ for
the first
two \eq s in the system (\ref{mot2}), which involve the variables $r,p_r$.
This point corresponds in the plane $\th=\pi/2$ to a unstable
circular orbit $\gamma$ of radius $r_u$. In this
case, we have also the presence of a $1-$parameter family of \hom
bounded orbits biasymptotic to $\gamma$ (but see also case {\em iii)}
below).

\sk\ni
{\em ii)} the degenerate situation where the unstable equilibrium point is
located at the infinity, i.e. $r=\infty,\ \.r=0$; the \hom orbits are
in this case a family of parabolas.

\sk\ni
{\em iii)} a ``critical'' case, with an unstable equilibrium point
$r_u\not=0$ and another unstable equilibrium point located at $r=\infty$,
and a family of heteroclinic orbits connecting these points.

\medskip

It is clear that -- due to the spherical \sy\ of
the problem -- all conclusions and properties stated for the plane
$\th=\pi/2$ are equally true for any plane for the origin in $\R^3$.

Let us now choose and single out the following \hom (heteroclinic in
case {\em iii)}) orbit in the plane $\th=\pi/2$, denoted by $\^\chi(t)$,
written in the spherical variables
\[
u\id(r,p_r,\phi,p_\phi,\th,p_\th)
\]
as follows
\beq
\^\chi(t)\id\big(R(t),\.R(t),\Phi(t),L, \pi/2,0\big) \lb{chi}
\eeq
where $R(t)$ and $\Phi(t)$ solve (\ref{mot2}) with the conditions
$R(\pm\infty)=r_u$ in case {\em i)} or respectively $R(\pm\infty)=\infty$
in case {\em ii)}, and
$R(-\infty)=r_u,\, R(+\infty)=+\infty$ in case {\em iii)}, and with
$\Phi(0)=\pi$. It is not necessary, for our purposes, to know explicitly
the expression of the functions $R(t)$ and $\Phi(t)$;  it will be useful
only to know that choosing $R(0)=r_0$ (the turning point) in cases {\em i)}
and {\em ii)}, $R(t)$ is an even function and $\Phi(t)$ an odd
function of the time $t$. Let us remark that any other \hom orbit can be
transformed by means of a rotation into the $\^\chi(t)$ given by (\ref{chi}).

We now introduce a smooth (analytic) perturbation depending in
general on all the variables $u$ and  time-periodic; the \eq s of
the motion we are considering are then, in general,
\begin{eqnarray}
\.r&=&p_r+\ep g_r(u,t) \nonumber\\
\.p_r&=&{p_\th^2\ov{r^3}}+{p_\phi^2\ov{r^3\sin^2\th}}
        -{dV\ov dr}+\ep g_{p_r}(u,t) \nonumber \\
\.\phi&=&{p_\phi\ov{r^2\sin^2\th}}+\ep g_\phi(u,t)\bigskip \lb{mp}\\
\medskip
\.p_\phi&=&\ep g_{p_\phi}(u,t) \nonumber\bigskip \\
\medskip
\.\th&=&{p_\th\ov r^2}+\ep g_\th(u,t) \nonumber\\
\.p_\th&=&{p_\phi^2\cos\th\ov{r^2\sin^3\th}}+\ep g_{p_\th}(u,t)\nonumber
\end{eqnarray}
where $\ep\ll 1$. Let us also write (\ref{mp}) in a more convenient
compact form
\beq
\.u=f(u)+\ep g(u,t)\lb{mpc}
\eeq
where
\beq f=J\grad_u H \lb{fJ} \eeq
and $J$ is the standard symplectic matrix.
Given any \hom orbit $\chi(t)$ of the unperturbed problem, one has that
$\chi_{t_0}\id\chi(t-t_0)$ satisfies, for all $t_0\in\R$,
\beq
{d\chi_{t_0}\ov dt}=f(\chi_{t_0}) \lb{fchi} \eeq

\medskip
In order to find conditions ensuring the occurrence of intersections
of stable and unstable manifolds of the critical point for the perturbed
problem, and hence the appearance of chaotic behaviour, we follow a
(suitably extended) procedure which has been first used in this context
(to the best of our knowledge) by Chow, Hale and Mallet-Paret \cite{CHM}
in a problem with $1$ degree of freedom.
We have first to look for smooth solutions near the homoclinic orbit;
we then put
\beq u(t)=\chi(t-t_0)+v(t-t_0)\lb{chiv}\eeq
Substituting in (\ref{mpc}), we find for $v(t)$ the equation (with the
time shift $t\to t+t_0$)
\begin{eqnarray}
\.v&=&A(t)v+\ep g(\chi(t),t+t_0)+ {\rm higher\ order\ terms\ in}\ v\
{\rm and}\ \ep \lb{ve0} \\
&\id &A(t)v+G(t,t_0,\ep) \nonumber
\end{eqnarray}
having used the shorthand notation $G(t,t_0,\ep)$, and where
\beq A(t)\id(\grad_u f)(\chi(t)) \lb{A}\eeq
This \eq , or more often its first-order approximation
\beq
\.v=A(t)v+\ep g(\chi(t),t+t_0)   \lb{ve1} \eeq
is called the variational \eq\ associated to (\ref{mpc}) and to the orbit
$\chi(t)$.

All the solutions $v(t)$ of (\ref{ve0}) can be written, in implicit form
(see \cite{Gr,CL})
\beq
v(t)=v_h(t) + \Psi(t)\int_{t_1}^t \Psi^{-1}(s)\, G(s,t_0,\ep)\, d s
\lb{vps}\eeq
where $t_1$ is arbitrary, $v_h(t)$ is any solution of the homogeneous
linear problem
\beq \.v_h=A(t)v_h \lb{veh} \eeq
and $\Psi$ is a fundamental matrix of solutions of (\ref{veh}). We now
have to look for the solutions of the homogeneous equation
(\ref{veh}) and to control their behaviour for $t\to\pm\infty$.
As well known \cite{Gr,CS2,CL}, this \eq\ admits some solutions
which remain bounded for all
$t\in\R$ and other solutions which diverge for $t\to\pm\infty$. The
asymptotic behaviour is exponential in the case of hyperbolic unstable
points, is like some power $|t|^\mu$ in the case of unstable point at
$r=\infty,\ \.r=0$, and more in general of nonhyperbolic critical points.

An obvious bounded solution is given by $d\chi/dt$. To find other
solutions, the following results, which we will state for convenience in
a quite general form, will be useful.

\bigskip

\section{General Results. The Role of Symmetry}
Let us start with the following definition (cf. \cite{Ol,CG})
\begin{definition}
A vector field
\beq
S\id\sigma(u)\cdot\grad =\sum_{i=1}^n \sigma_i(u){\pd\ov{\pd
u_i}} \lb{LPS} \eeq
is said to be a Lie-point \sy\ (or -- more exactly -- the Lie generator
of a Lie-point \sy ) for a general dynamical system (not necessarily \Ham )
\beq \.u=f(u) \lb{DS} \eeq
if
\beq [\sigma\cdot\grad , f\cdot\grad]=0 \lb{Lp} \eeq
where $[\cdot , \cdot]$ denotes the usual Lie commutator.
\end{definition}
We then can state:
\begin{lemma}
Let $\chi(t)$ be any given \hom orbit, solution of the problem (\ref{DS}),
and assume that this problem admits a Lie-point \sy\
$S=\sigma(u)\cdot\grad$. Then
\[ \zeta(t)\id \sigma(\chi(t)) \]
is a solution of the homogeneous part (\ref{veh})  of the
variational  \eq\ associated to this problem.
\end{lemma}
\noindent
{\em Proof}. From direct calculation, using (\ref{Lp}) and (\ref{A}),
\[
\.\zeta={d\ov dt}\sigma(\chi(t))=\grad_u\sigma\cdot\.\chi=
(f\cdot\grad_u\sigma)(\chi(t))=(\grad_u f\cdot\sigma)(\chi(t))=A(t)\zeta.
\]

\sk\ni
{\bf Remark}. The obvious solution $d\chi/dt$ of (\ref{veh}) may be
included into the solutions given by the above Lemma, indeed
$f\cdot\grad=d/dt$ is a (trivial) Lie-point \sy\ of any autonomous
dynamical system (\ref{DS}), expressing simply its time-flow invariance:
if $u_0(t)$ is a solution, then $u_0(t+t_0)$ is also a solution.

\noindent
\begin{lemma}
Given an \Ham\ $H$, let $K$ be a constant of motion for the \Ham ,
i.e. $\{K,H\}=0$. Then
\beq
S\id J\grad K\cdot\grad \lb{SK} \eeq
is a Lie-point \sy\ for the dynamical system $\.u=J\grad_uH$.
\end{lemma}
The proof is a straightforward verification.

\bigskip

\section{General Results. \\ Mel'nikov-type Conditions}
Still considering, from a general point of view, eq. (\ref{mpc}) as arising
from an \Ham\ with $n$ degrees of freedom, assume that there are $n$ bounded
solutions $\zeta^{[\a]}(t)$ and $n$ divergent solutions $\eta^{[\a]}(t)$
of the
homogeneous part (\ref{veh}) of the variational \eq . Let us construct
the fundamental matrix $\Psi$ in (\ref{vps})
putting the solutions $\zeta^{[\a]}$ in the first $n$ columns and
the $\eta^{[\a]}$ in the remaining columns of $\Psi$; recalling that
$\det\Psi=1$ and observing that $\Psi^{-1}$ is a matrix having  the
$\eta^{[\a]}$ in the first rows and $-\zeta^{[\a]}$ in the remaining rows,
one easily sees (cf. \cite{Gr}) that (\ref{vps}) can be written
(still in implicit form, see (\ref{ve0}))
\begin{eqnarray}
v(t)=v_h(t)+&\sum_\a &\zeta^{[\a]}(t)\int_{t_1}^t\Big(\eta^{[\a]}(s)\cdot
J\, G(s,t_0,\ep)\Big)\, ds \lb{sove} \\
&+&\eta^{[\a]}(t)\int_{t_1}^t\Big(\zeta^{[\a]}(s)\cdot J\,
G(s,t_0,\ep)\Big)\, ds \nonumber
\end{eqnarray}
where $v_h(t)$ is a linear combination of the
$\zeta^{[\a]}(t),\eta^{[\a]}(t)$.

We now have to look for the existence of bounded solutions of
(\ref{ve0}); more precisely we have to look for solutions
$v^{(-)}(t)$ \big(and resp. $v^{(+)}(t)$\big)
with the property of being bounded for $t\to-\infty$ (resp. $t\to+\infty$);
these will provide precisely those solutions
\beq
u^{(\pm)}(t)=\chi(t-t_0) +v^{(\pm)}(t-t_0) \lb{chpm} \eeq
of (\ref{mpc}) which belong, by definition, to the unstable (resp. stable)
manifold of the critical point.

If we linearize (\ref{sove}) around the solution $v(t)\equiv 0$ (which
amounts to deleting the higher-order terms in (\ref{ve0}), or to
considering the variational equation in the form
(\ref{ve1})), and take into account the different asymptotic behaviour of
the solutions $\zeta^{[\a]}(t)$ and $\eta^{[\a]}(t)$, it can be seen,
as a consequence of the implicit-function theorem \cite{CHM,Gr,CS2},
that the existence
of bounded solutions {\em both} for $t\to -\infty$ and $t\to+\infty$ is
ensured if the following Mel'nikov-type conditions are verified
\beq
M^{[\a]}(t_0)=\intR \Big(\zeta^{[\a]}(t) \cdot J\, g(\chi(t),t+t_0)\Big)\,
dt=0 \lb{M}
\eeq
Let us remark that if the perturbation is \Ham , $g=J\grad_uW$, as often
happens in celestial mechanics, choosing $\zeta^{[1]}(t)=d\chi/dt$, then
the first of these conditions becomes
\beq M^{[1]}(t_0)=\intR \{H,W\}(\chi(t),t+t_0)\, dt\ =\ 0 \lb{M1H}\eeq
Similarly, if, e.g., $\zeta^{[2]}(t)$ comes from a constant of motion $K$,
according to Sect. 3, then
\beq M^{[2]}(t_0)=\intR \{K,W\}(\chi(t),t+t_0)\, dt\ =\ 0 \lb{M2H} \eeq
These conditions are identical to the conditions given e.g. in
\cite{W,HM,Rc},
where they are obtained by means of different procedures and hypotheses.
It can be significant to remark that the present method then provides an
extension of these formulas also to
``degenerate'' cases (nonhyperbolic points and possibly unbounded
orbits), and to \sys\ of more general nature, as in (\ref{LPS}).

We have only to notice that, whereas the convergence of Mel'nikov
integrals (\ref{M}) is granted in the case of standard hyperbolic
and isolated unstable
points, the convergence is only ``conditional'', i.e. along a suitable
sequence of intervals (see \cite{W} for any details), in cases
{\em i)} and {\em iii)} of our classification
in Sect.~2,  and finally it must be controlled ``by hand''
in the non-hyperbolic case or in the case
of critical point at the infinity. In the last cases one has to impose a
sufficiently rapid vanishing of the contribution  of the perturbation
$g(\chi(t),t)$ when $t\to\pm\infty$, i.e., as expected, when approaching
the critical point. The precise rate of this
vanishing will depend on the specific problem in consideration (see
Examples in the next sections, and also \cite{CS1} for examples in $1$
degree of freedom).  For what concerns the regularity of the solutions
and of the asymptotic manifolds, see e.g. \cite{DH,Rj,CFN}.

Changing now the point of view, and considering the Poincar\'e sections of
the $u^{(-)}$ and $u^{(+)}$ solutions, the above arguments show that,
once conditions (\ref{M}) are satisfied, there occurs a crossing of the
negatively and positively asymptotic sets on the Poincar\'e section
\cite{GH,W,HM}. One
usually imposes that the intersection is transversal; actually this
condition is not strictly necessary, indeed it can be shown that
it is sufficient that the crossing is ``topological'' \cite{BW}, i.e.,
roughly, that there is really a ``crossing'', from one side to the other,
but we do not insist on
this point, which goes beyond the scope of the present paper.

Thanks to the periodicity of the perturbation, one immediately
deduces \cite{GH,W,CHM,BW} that there is an infinite sequence of
intersections of the positively and negatively asymptotic sets of the
critical point in the Poincar\'e section, leading to a situation typical
of the \hom chaos.
The presence of such infinitely many intersections is clearly reminiscent
of the chaotic behaviour expressed by the Birkhoff-Smale theorem in terms
of the equivalence to the symbolic dynamics of Smale
horseshoes. Actually, this theorem
cannot be directly used in the present context because its standard
proof is intrinsically based on hyperbolicity properties \cite{GH,W}.
However, several arguments can be invoked even in the
``degenerate'' cases, which allow us to conclude that, if the conditions
(\ref{M}) are satisfied, the perturbed problem exhibits a chaotic behaviour.
We can refer e.g. to the classical arguments used in \cite{Mo}, and
reconsidered
by many others (see e.g. \cite{LM,Geh,DSe,X1}), possibly resorting to
singular coordinate transformations, such as the McGehee transformation
in the case of critical point at the infinity, or the ``blowing-up''
method \cite{Du,BM}. More specifically, an
equivalence to a ``nonhyperbolic horseshoe'' has been proved in \cite{DH},
in which
the contracting and expanding actions are not exponential but
``polynomial'' in time. The presence
of Smale horseshoes and of a positive topological entropy has been also
proved by means of a quite general geometrical or
``topological'' procedure \cite{BW} which holds, in the presence of
area-preserving perturbations, even in the
nonhyperbolic case.

\bigskip

\section{Applications to Celestial Mechanics.\\
Two Degrees of Freedom}
Coming back to the initial problems as stated in Sect.~2, the first step,
according to the above discussion, is to write down the appropriate
Mel'nikov conditions. It is clear that, apart from a rotation (this
will change the expression of the perturbation $g$ in equations
(\ref{mp},\ref{mpc}) into some new $\~g$ which will
also depend in general on the parameters of the rotation: this will be
explained in detail later), we can always assume that the \hom orbit lies
in the plane $z=0$ and in particular is given precisely by the orbit
$\^\chi(t)$ defined in (\ref{chi}), Sect.~2. This greatly simplifies
the variational equation, and in particular its homogeneous part
(\ref{veh});
indeed it is immediate to verify that, with  this choice, the two \eq s
for the variations $v_5,v_6$ of $\th$ and $p_\th$ are separated from
the first four \eq s and admit regular bounded \so s; therefore the
problem turns out to be 4-dimensional. An obvious \sy\ for it is given by
the rotations around the $z-$axis, generated by
\beq S={\pd\ov{\pd \phi}} \lb{rot} \eeq
Then, as discussed in Sect.~3, two bounded \so s of the homogeneous part
of the variational \eq\ are
\begin{eqnarray}
\zeta^{[1]}&=&{d\^\chi\ov dt}=(\.R(t),\"R(t),\.\Phi(t),0,0,0) \lb{ze} \\
\zeta^{[2]}&=&(0,0,1,0,0,0)\nonumber
\end{eqnarray}
where $R(t),\, \Phi(t)$ have been defined in Sect. 2 (see
(\ref{mot2},\ref{chi})). We then get from (\ref{M}) the two Mel'nikov
conditions
\begin{eqnarray}
M^{[1]}&=&\intR\Big[
\.R(t)\~g_{p_r}(\^\chi(t),t+t_0)-\"R(t)\~g_{r}(\ldots)  \lb{M1} \\
&\q& \qq\q +\ \.\Phi(t)\~g_{p_\phi}(\ldots)\Big] dt \ =\ 0 \nonumber\\
M^{[2]}&=&\intR\~g_{p_\phi}(\^\chi(t),t+t_0)\ dt \ =\ 0 \lb{M2}
\end{eqnarray}
Notice that usually one has $\~g_{r}=0$. In the case  where the
problem, included the
perturbation, is completely planar, and that the perturbation is
generated by an \Ham\ $W$ of the form
\beq W=W(r,\phi,t) \lb{Wrft} \eeq
then, under rotation, $W$ is changed simply into $W(r,\phi+\phi_0,t)$ and
the above conditions become
\beq
M^{[1]}(t_0,\phi_0)=\intR\Big[ \.R(t){\pd W(R(t),\Phi(t)+\phi_0,t+t_0)\ov \pd
r}+\.\Phi(t){\pd W(\ldots)\ov \pd \phi} \Big] dt\ =\ 0 \lb{M1W}
\eeq
and
\beq
M^{[2]}(t_0,\phi_0)=\intR {\pd W(R(t),\Phi(t)+\phi_0,t+t_0)\ov \pd\phi}
\, dt\ =\ 0 \lb{M2W}
\eeq
and it is easily seen that these can be also written according to
the general form as given in (\ref{M1H},\ref{M2H}).

As a first simple application of this situation, consider the classical
Kepler problem with unstable equilibrium point at $r=\infty$ and $\.r=0$
and a perturbation not depending on $\phi$, e.g.
\beq W={\sin 2\pi\nu t\ov r^\de}\qq\qq (\de>1/2) \lb{Gy} \eeq
as in the classical Gyld\'en problem \cite{CS1,DSe}. Then,
condition (\ref{M2W}) is
automatically satisfied, and the integral in (\ref{M1W}) becomes
(thanks to the parity of the function $R(t)$)
\[
\cos 2\pi\nu t_0\intR{\.R(t)\ov {R(t)^{\de +1}}}\sin 2\pi\nu t \ dt
\]
which clearly admits simple zeroes (the integral is easily seen to be
$\not=0$). In this example, it is also simple to evaluate explicitly the
asymptotic behaviour of the \so s of the variational \eq : one has indeed
\cite{CS2}
\beq \zeta=\.R(t)\sim |t|^{-1/3}\q , \q \eta(t)\sim |t|^{4/3} \qq {\rm
for}\q |t|\to +\infty
\eeq
and all conditions for the procedure given in Sect. 4 are satisfied, with
-- in particular -- the condition $\de>1/2$ which ensures in this case
the correct rate of vanishing of the perturbation at $r=\infty$. Notice
that this same condition would also guarantee that under
the McGehee transformation \cite{Mo,Geh,DSe}, the perturbation is not
singular. Then chaotic behaviour is expected for this problem.

A similar result holds for the more general (time-periodic) perturbations
of the form (\ref{Wrft}) occurring e.g. in the restricted 3-body problems
\cite{Rc,X1}.
These cases can be conveniently dealt with in this way.
Assuming that $W(R(t),\Phi(t),t)\to~0$ for
$t\to\pm\infty$, then (\ref{M1W}) (or (\ref{M1H})) becomes
\beq M^{[1]}(t_0,\phi_0)=\intR {\pd W\ov \pd t}(R(t),
\Phi(t)+\phi_0,t+t_0)\, dt\ = \ 0 \lb{MD}\eeq
(where clearly the derivative $\pd/\pd t$ must be performed only with
respect to the explicit time-dependence of $W$);
introducing then the ``Mel'nikov potential'' ${\cal W}={\cal W}(\phi_0,t_0)$
(cf. \cite{Del}), corresponding to the perturbation $W(r,\phi,t)$:
\beq
{\cal W}(t_0,\phi_0)\id\intR W\big(R(t),\Phi(t)+\phi_0,t+t_0\big)\, d t
\lb{MP}\eeq
then one gets from this definition and from (\ref{MD},\ref{M2W})
\beq
M^{[1]}(t_0,\phi_0)={\pd {\cal W}\ov{\pd t_0}}=0 \qq\ , \qq
M^{[2]}(t_0,\phi_0)={\pd {\cal W}\ov{\pd \phi_0}}=0 \lb{M12}\eeq
In other words, the two Mel'nikov conditions are equivalent to the
existence of stationary points  for the Mel'nikov potential
${\cal W}(\phi_0,t_0)$. On the other hand, ${\cal W}$ is a smooth
doubly-periodic
function, and such a function certainly possesses  points $\bar{t_0},
\bar{\phi_0}$ where the two partial derivatives in (\ref{M12}) vanish, and
this implies that the two conditions (\ref{MD},\ref{M2W}) are certainly
satisfied (one has only to check that these stationary points of
${\cal W}$ are isolated).

\bigskip

\section{Applications to Celestial Mechanics. \\
Three Degrees of Freedom}
Let us consider finally a perturbation depending on both angles
$\phi,\theta$. As already stated, to obtain Mel'nikov condition in the
form (\ref{M1},\ref{M2}), one has to transform the generic \hom orbit
into the orbit $\^\chi(t)$ given by (\ref{chi}). This is obtained by
means of a rotation defined by the
following Euler angles (with the conventions and notations as in
\cite{Go}):
\beq -\Omega,\ - i,\ - \om
\eeq
where, with the language of celestial mechanics, $i$ is the inclination
of the plane of the orbit, $\om$ the angle of the perihelion with
the line of nodes
in the orbital plane, and $\Omega$ is the longitude of the ascending node.

Considering for simplicity a perturbation generated by an \Ham\ not
depending on the variables $p$, which we denote here, with a little
abuse of notation, either by $W(r,\phi,\th,t)$ or by
$W({\bf x},t),\, {\bf x}=(x,y,z)$,
and denoting by $B$
the matrix of this rotation, the new expression $\~W$ of the perturbation
is obtained by replacing ${\bf x}$ with $B{\bf x}$. It is then easy to
verify that Mel'nikov conditions (\ref{M1},\ref{M2}) become
\begin{eqnarray}
M^{[1]}(t_0,\om,\Omega)&=&\intR\Big[\.R{\pd W\ov{\pd r}}+
\.\Phi\Big({\pd\~W\ov{\pd \phi}}\Big)_0\Big]\, dt =0 \lb{Mt1} \\
M^{[2]}(t_0,\om,\Omega)&=&\intR\Big({\pd\~W\ov{\pd \phi}}\Big)_0\, dt =0
\lb{Mt2}
\end{eqnarray}
where
\beq
\Big({\pd\~W\ov{\pd\phi}}\Big)_0=
  R(t)\Big(- C_1\big(R(t),\Phi(t),\om,\Omega\big)\, \sin \Phi(t)+
C_2(\ldots)\cos\Phi(t)\Big)
\eeq
with
\begin{eqnarray}
C_1&=&\Big({\pd \~W\over{\pd x_1}}\Big)_0(\cos\om\cos\Omega-
\cos i \sin\om\sin\Omega)+ \nonumber \bigskip\\
&+&\Big({\pd \~W\over{\pd x_2}}\Big)_0
(-\cos\om\sin\Omega-\cos i\sin\om\cos\Omega)+
\Big({\pd \~W\over{\pd x_3}}\Big)_0\sin i\sin\om  \nonumber\\
\medskip
C_2&=&\Big({\pd \~W\over{\pd x_1}}\Big)_0(\sin\om\cos\Omega+
\cos i \cos\om\sin\Omega)+ \nonumber \sk\\
&+&\Big({\pd \~W\over{\pd x_2}}\Big)_0
(-\sin\om\sin\Omega+\cos i\cos\om\cos\Omega)-
\Big({\pd \~W\over{\pd x_3}}\Big)_0\sin i\cos\om \nonumber
\end{eqnarray}
and where $(\pd \~W/\pd x_i)_0$ means that in the derivative of the given
$W$ with respect to $x_i$ one has to replace ${\bf x}$ with $B{\bf x}$ and
finally put $z=0$ (or $\theta=\pi/2)$.

If $i=0$, i.e. if the
problem is completely planar, including the perturbation, or if the
perturbation is ``generic'', i.e. has no ``preferred'' direction in the
space (as often happens, see below for an example), and therefore it is not
restrictive to choose $i=0$, then one gets
\[
\q C_1=\Big({\pd \~W\ov \pd x_1}\Big)_0\cos\phi_0 +
\Big({\pd \~W\ov \pd x_2}\Big)_0\sin\phi_0\ , \q
C_2=\Big({\pd \~W\ov \pd x_1}\Big)_0\sin\phi_0 -
\Big({\pd \~W\ov \pd x_2}\Big)_0\cos\phi_0
\]
where $\phi_0=-(\om+\Omega)$, and the above expressions
(\ref{M1W},\ref{M2W}) are recovered.

An example is provided by a problem in general relativity \cite{LV,SC}.
Consider indeed the motion of a relativistic charged particle in a
gravitational field produced by a Schwarzschild black hole, perturbed by a
homogeneous constant electric field. It can be shown \cite{SC} that the
perturbation is given by
\beq W={{\cal F}}(r)(l_1 \cos\phi\sin\theta+ l_2
\sin\phi \sin\theta +l_3 \cos\theta) \eeq
where ${{\cal F}}(r)$ is suitable function, and
${\bf l}=(l_1,l_2,l_3)$ is the direction of the perturbing field.
This direction is generic and therefore no rotation is required (it
would simply change the values of $l_i$, which are not fixed).

In this case, the first Mel'nikov condition in (\ref{Mt1}) is identically
satisfied, because the perturbation is independent of the time
(cf. (\ref{MD})), and the other one becomes
\beq M^{[2]}=M^{[2]}(\phi_0)=
(l_1\sin\phi_0 -l_2\cos\phi_0) \intR {{\cal F}}(R(t)) \cos\Phi(t)~ dt \eeq
which clearly admits simple zeroes, if only one can show that the
integral is different from zero (see \cite{LV}).

Notice that in this example one has that the effective unperturbed
potential  admits bounded \hom orbits biasymptotic to an unstable point
$r_u\not=0$ (which corresponds to unstable circles, as in case {\em i)}
of our initial classification), therefore
the above integral is expected to converge only conditionally \cite{W,Rc}.

Similar results hold if the perturbation is produced by a constant
homogeneous magnetic field \cite{SC}.
It can be observed that, whereas the component of  the electric
and of the magnetic  field  on the plane of motion  leads to a chaotic
dynamics, the component normal to the plane does not.
Since the problem is spherically symmetric,  this argument can be applied
to every plane for the origin. Thus,  given an electric or magnetic
field,  on each  plane for the origin (except at most  the one normal to
the field),  chaos appears for a suitable choice of initial
conditions.


\end{document}